\documentclass{article}

\usepackage{arxiv}

\usepackage[utf8]{inputenc} 
\usepackage[T1]{fontenc}    
\usepackage{hyperref}       
\usepackage{url}            
\usepackage{booktabs}       
\usepackage{amsfonts}       
\usepackage{nicefrac}       
\usepackage{microtype}      
\usepackage{lipsum}
\usepackage{graphicx}
\usepackage{nameref}
\usepackage{makecell}
\usepackage{appendix}

\title{Design Knowledge Representation with\\ Technology Semantic Network}

\author{
  Serhad Sarica \\
  Data-Driven Innovation Lab\\
  Singapore University of Technology and Design\\
  Singapore, 487372 \\
  \texttt{serhad\_sarica@mymail.sutd.edu.sg} \\
   \And
 Jianxi Luo \\
  Data-Driven Innovation Lab\\
  Singapore University of Technology and Design\\
  Singapore, 487372 \\
  \texttt{luo@sutd.edu.sg} \\
}

\begin{document}
\maketitle

\begin{abstract}
Engineers often need to discover and learn designs from unfamiliar domains for inspiration or other particular uses. However, the complexity of the technical design descriptions and the unfamiliarity to the domain make it hard for engineers to comprehend the function, behavior, and structure of a design. To help engineers quickly understand a complex technical design description new to them, one approach is to represent it as a network graph of the design-related entities and their relations as an abstract summary of the design. While graph or network visualizations are widely adopted in the engineering design literature, the challenge remains in retrieving the design entities and deriving their relations. In this paper, we propose a network mapping method that is powered by Technology Semantic Network (TechNet). Through a case study, we showcase how TechNet's unique characteristic of being trained on a large technology-related data source advantages itself over common-sense knowledge bases, such as WordNet and ConceptNet, for design knowledge representation.
\end{abstract}


\section{Introduction}
\label{sec:intro}

Automatic summary representation of design-related topics or entities in technical design documents is an important task in engineering design since it can inform designers in various tasks in different phases of the design process  \cite{Dong1996, Szykman2000}. For instance, engineering design researchers have studied the topics in large design repositories to reveal the prominent and emerging fields \cite{Chiarello2019}, or to discover the structure of these repositories and enable the search for prior arts and design inspiration (or stimuli) in the early design stages \cite{Fu2013, Song2020}. Topic mapping methods can provide various insights, such as most frequently addressed topics or particular topics within a collection of documents \cite{Rehurek2010}.

However, training on a local document dataset and associating rules might not lead to a model representing the true technical associations of the design concepts appearing in the document(s). A specific concept that is fundamentally essential for a design may not appear statistically significant in a single document or a set of documents that describe the design. One can read the document and try to identify different concepts and how they are related if he/she is knowledgeable in the field of the corresponding design. Nevertheless, such a human reading process is tedious, time-consuming, and prone to human cognitive errors or limitations. Alternatively, knowledge bases such as domain-specific ontologies \cite{Li2005}, WordNet \cite{Fellbaum2012, Miller1990}, and ConceptNet \cite{Speer2012, Speer2017a, Speer2017b} can be used to retrieve the design entities from a design document and determine their relations to provide a structured representation of the design.

Among these knowledge bases, the semantic network of TechNet \cite{Sarica2020} is derived from a vector space of technical terms that are statistically derived from the complete patent database. Thus, it has a unique characteristic of drawing technically sound relations between the technical terms universally. This characteristic provides TechNet an edge in the data-driven and artificial intelligence oriented downstream engineering design tasks comparing to widely recognized common-sense knowledge bases, such as ConceptNet and WordNet. The interest of this study is to automatically and visually represent a technical text to summarize and map the entities of the design described by the text without the need to read the long and complex technical description.

A graph or network of the terms within the technical text may be considered as one of the visual representation methods, such as word-clouds, bubble-graphs, occlusion, etc\footnote{A great number of examples can be found in https://observablehq.com/@d3/gallery.} , to present an overall summary of design entities and their relations in the design. One advantage of graph representations is that it allows the application of network layout methods that can position closely related terms (the terms linked by a high-weighted edge) together. Therefore, it can clump the semantically relevant terms together in a positionally distinct and more confined space in the visualization. Hence, we utilized networks in our method to visualize and evaluate the design information representation power of the TechNet compared to the benchmark knowledge bases, namely ConceptNet and WordNet. 

The next sections review the related studies in the engineering design literature briefly, present a generic methodology to create a graph visualization of a technical text using a publicly available knowledge base, and demonstrate the power of TechNet in providing meaningful graph representation of a technical text using a comparative case study.

\section{Related Work}
\label{sec:approach}
Traditional topic mapping and modelling methods reveal the common topics in a set of documents and latent semantic structures in the documents by employing mainly Latent Semantic Analysis (LSA) \cite{Deerwester1990} or Latent Dirichlet Allocation (LDA) \cite{Blei2003}. These topic modelling methods have been extensively employed in the engineering design literature to to create structured design repositories \cite{Fu2013}, aid prior art or document search \cite{Krestel2013}, enable analysis of longitudinal changes in a field \cite{Chiarello2019}, and support the innovative product design processes \cite{Song2020, Dong2004}. Studies using traditional topic modelling methods provide more coarse information about documents and their contents, which can be used to map them to meaningful groups in a set of documents. These studies do not focus on detailed design aspects noted in these documents.

Summarizing and representing the design-related technical information within a design document or technical description requires identifying the technical concepts and the technically-meaningful relations among them. Knowledge bases, such as lexical databases, semantic networks, and knowledge graphs, may provide information about entities and their relations. However, they are limited by their coverage and ability to represent the relations as close as possible to reality. Several studies offered methods of constructing ontologies for specific engineering design domains \cite{Ahmed2007, Gero2014, Li2007}, such as the Function-Behavior-Structure (FBS) ontology, which proposes a structured and standardized way of representing design and designing \cite{Gero2014}. However, these studies lack a generalizable method of populating these ontologies beyond the domains of interest without the involvement of human experts. Therefore, their usage in domains different from those they model is not feasible and realistic.

Larger knowledge bases, such as the largest manually created lexical database, namely WordNet \cite{Fellbaum2012, Miller1990}, and common-sense semantic network, ConceptNet \cite{Speer2012, Speer2017a, Speer2017b}, may provide an alternative solution platform for retrieving the technical entities and relations among them in a design description. However, being common-sense knowledge bases, they primarily focus on layman knowledge, and detailed technical terms may be overlooked while constructing these knowledge bases and the term-to-term associations are not contextualized in engineering \cite{Sarica2020}.

On the other hand, engineering and technology-related larger knowledge bases, such as B-link \cite{Shi2017} and TechNet \cite{Sarica2020}, address the lack of representation of technical terms and their technical associations. B-link retrieved technical terms from engineering-related academic papers and design-related online web platforms, and derived relations among them using co-occurrence information. TechNet used patents as the data source, created a collection of over 4 million terms, and trained language models to represent those terms with continuous word embeddings. Both semantic networks have shown that their coverage of technical terms is superior to other competitive knowledge bases. In addition, TechNet is superior in relating technical terms with respect to engineers' comprehension, and it makes all the information available via an interface and public APIs\footnote{TechNet interface is accessible via http://www.tech-net.org/ and API definitions are documented in TechNet's GitHub repository https://github.com/SerhadS/TechNet}. 

The graph- or network-based methods are vastly used in the engineering design literature to represent the relatedness structure of design entities or documents in various tasks. Network metrics have provided a medium to derive useful design-related insights from the structure of the graphs, and various layout methods have provided ways of representing the design-related data in an easily comprehensible way \cite{Song2018, Lim2016}. For example, network visualizations have been utilized to represent the whole technology space to support innovation and competitive intelligence \cite{Luo2017, Luo2019, Sarica2020b}, show the relations between components and subsystems to evalute designs \cite{He2017, Pasqual2012, Sosa2007} and inform design decisions \cite{Song2018, Sosa2007, Kim2012}, discover the patterns of design activities \cite{Cash2014, Cash2015}, reveal the structure of design document repositories to guide retrievals \cite{Fu2013}, and represent mind maps \cite{Camburn2020b, Camburn2020} and concept networks \cite{Shi2017, Chen2020, Liu2020, Sarica2019, Song2020b, Souili2015} for design ideation uses. On the other hand, a few studies explored other visualization methods such as word-clouds \cite{He2019, He2019b} based on design description texts.

Although one can rigorously read and study a design document or description to discover all the design-related entities and comprehend their relations, such a human process is tedious, labour-intensive, and limited by the domain-specific knowledge of the reader. Knowledge bases such as semantic networks or knowledge graphs may provide the necessary infrastructure, a pre-trained database of semantic entities and their semantic associations, to retrieve the design-related technical terms and relations between them, automatically. Our work focuses on providing a generic methodology to visually and succinctly represent and summarize design specific entities in technical language descriptions, based on a technology-related semantic network, namely TechNet. Specifically, we assessed the effectiveness of TechNet in providing a meaningful representation of a technical description compared to other publicly available knowledge bases, namely ConceptNet and WordNet.

\section{Methodology}
\label{sec:methodology}

\begin{figure}[h]
  \centering
  \includegraphics[scale=0.50]{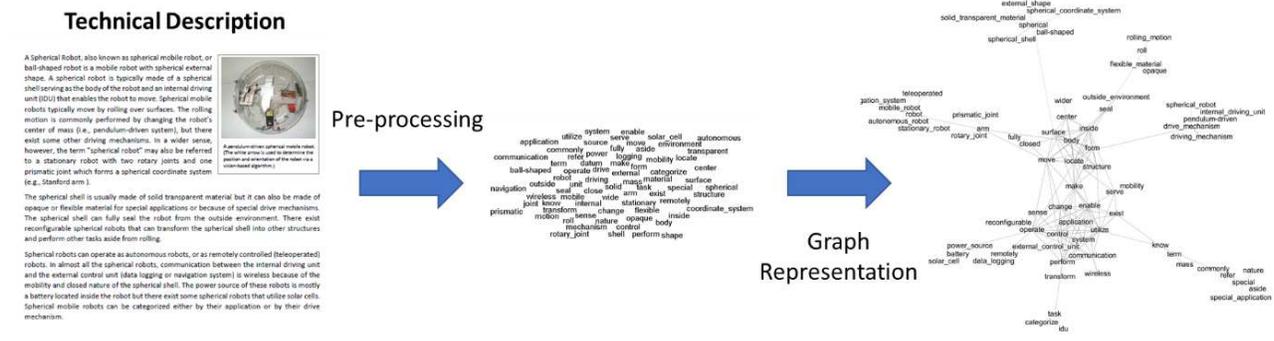}
  \caption{Overview of the methodology}
  \label{fig:fig1}
\end{figure}

We followed a generic procedure to create a graph representation of a technical text regardless of the knowledge base employed, as explained below and represented in Figure \ref{fig:fig1}.

\begin{enumerate}
\item Pre-processing step: Retrieve all the terms, including phrases from the text that are contained in the selected knowledge base. Since the knowledge bases have different vocabularies, the number and variety of the terms retrieved by different knowledge bases naturally vary.
\item Create an adjacency matrix $A$ with a size of $N$ where $N$ is the size of the term set found in step 1, and $A_{ij}$ corresponds to the semantic similarity of terms $i$ and $j$ in the corresponding knowledge base. $A_{ij}$ relates to path similarity of $i$ and $j$ for WordNet and cosine similarity of vector representations of $i$ and $j$ for ConceptNet and TechNet.
\item The edge filtering method offered by Hidalgo et al. \cite{Hidalgo2007}, Yan \& Luo \cite{Yan2017}, and Luo et al. \cite{Luo2017} is applied to produce a network visualization that can effectively represent the relational design knowledge within the text while reducing the complexity of the visualization to enable rapid comprehension within minutes. The edges are filtered to generate a maximum spanning tree (MST). An MST has the minimum set of the strongest edges in the graph to keep the graph connected in which $N-1$ edges between $N$ nodes are to be expected. Starting with the MST as the backbone of the term relations, we superposed the next strongest edges until we have $2N$ edges in the graph.
\item A force-directed algorithm \cite{Jacomy2014} is run on the resultant network graph to reach a stable layout.
\end{enumerate}

Having a generic method to visualize a graph representation of the text, we retrieve terms and their associations in three different knowledge databases: TechNet, WordNet, and TechNet. TechNet, ConceptNet, and WordNet are constructed with different methods and procedures, using different data sources. Thus, sets of terms they contain are different from each other. TechNet’s superior coverage of technical terms was already demonstrated in Sarica et al \cite{Sarica2020}.

All these three knowledge bases have the necessary structure and capabilities to quantitatively represent the semantic similarity between the terms they contain. In this study, we used the WordNet path similarity \cite{Pedersen2004} as the semantic similarity metric of WordNet, which returns a quantitative measure based on the shortest path that connects the terms. For TechNet and ConceptNet, we used cosine similarity of vector representations of terms in these knowledge bases. In Sarica et al. \cite{Sarica2020}, these three knowledge bases and others are compared according to their performances on retrieving semantic relations between technology-related terms according to engineers' comprehension, showing the superiority of TechNet over others.

Here we further hypothesize that TechNet is also superior for generating topic maps that represent the design information in technical texts in a more comprehensible, detailed, and structured way. We test the hypothesis in the case study below.

\section{Case Study: Spherical Rolling Robot Technical Knowledge Representation}
\label{sec:case}
\begin{figure}[h]
  \centering
  \includegraphics[scale=0.55]{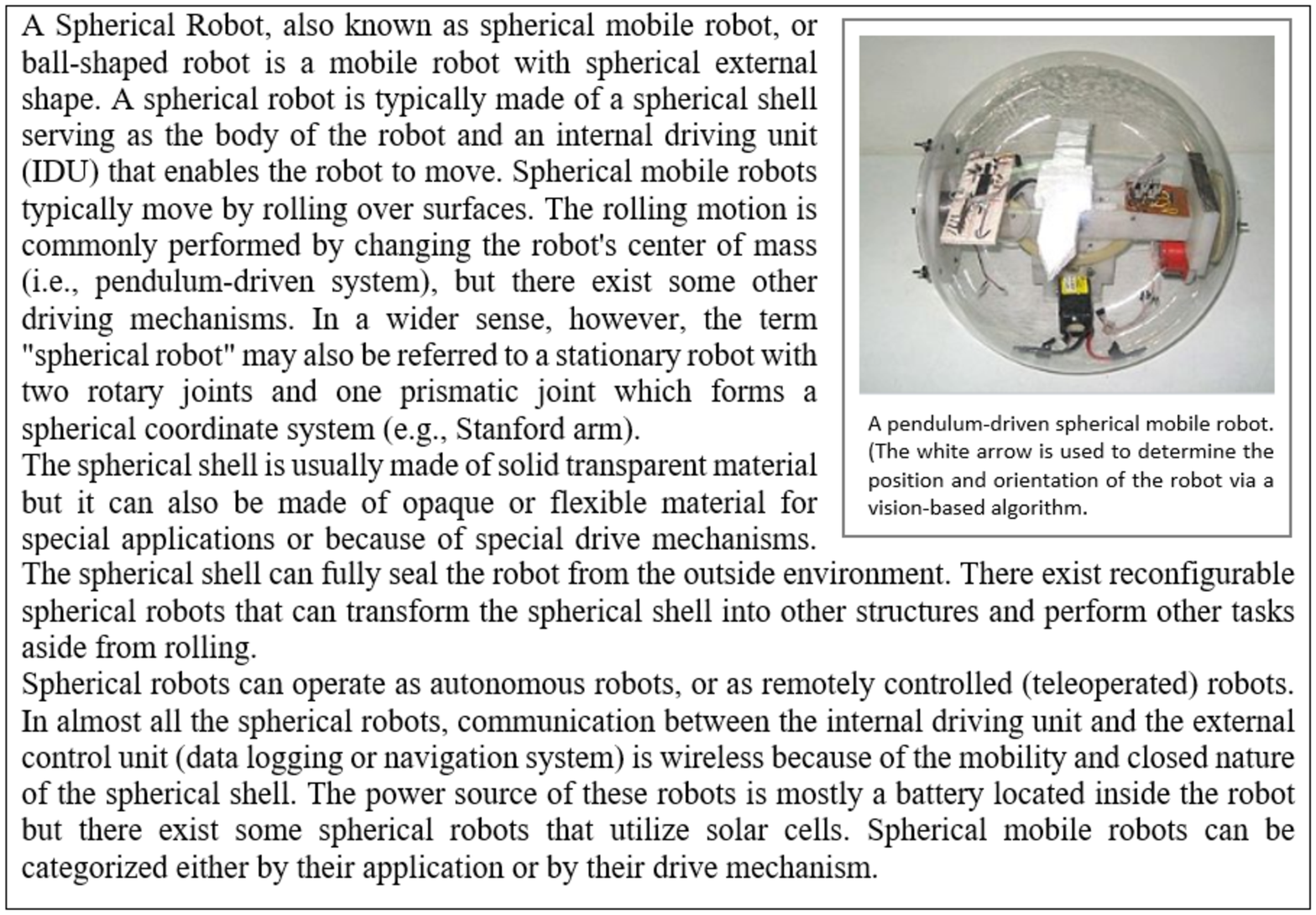}
  \caption{The technical description of “Spherical Robot” used in the case study (Source: Wikipedia, accessed on June 8, 2020)}
  \label{fig:fig2}
\end{figure}

A technical text can be a part of a document or a whole document such as a technical article, a scientific paper, a patent text, a technical description of a product, a maintenance or user manual, a system design review, or a technical definition of a technology that describes an entity using technical jargon. In this case study, first, a text description of a specific technology that typical engineers can potentially comprehend and relate to is selected (Figure \ref{fig:fig2}). The text specifically focuses on the design of “spherical robots”, which integrate technologies from different generic fields (mechanical and electronics). The spherical robot concept has also gained popularity with BB-8’s appearance in the Star Wars movies. Spherical robots are spherical shaped robots that generally roll to move on a surface and combine many mechanical and electronic components to achieve this rolling motion and enable possible capabilities such as remote controlling, autonomous abilities, and data collection with built-in sensors.

Following the methodology, the pre-processing step is conducted separately for each knowledge base to retrieve the terms from the technical description. The resultant term lists are different because of the significant difference in the vocabularies contained in these knowledge bases. As a result, 75, 77, and 73 unique terms are retrieved from the same text (in Figure \ref{fig:fig2}) based on the vocabularies of WordNet, ConceptNet, and TechNet, respectively. Table \ref{table:Tab1} presents the number of unigrams, bigrams, and trigrams retrieved from the same technical text using different knowledge bases. TechNet’s comprehensive technology-focused vocabulary enables retrieval of even long technology-related entities such as “internal driving unit”, “spherical coordinate system”, and “solid transparent material” and possesses a clear distinction from WordNet and ConceptNet.

\begin{table}[h]
\centering
\caption{Top 30 terms for term-frequency, IDF, TFIDF and entropy}
\begin{tabular}{ccccc}
\hline
            & Uni-gram  & Bi-gram   & Tri-gram  & Total    \\
\hline
WordNet     & 72        & 3         & -         & 75     \\
ConceptNet  & 72        & 5         & -         & 77    \\
TechNet     & 52        & 17        & 4         & 73        \\
\hline
\end{tabular}
\label{table:Tab1}
\end{table}

\begin{figure}[!htb]
	\centering
	\includegraphics[scale=0.81]{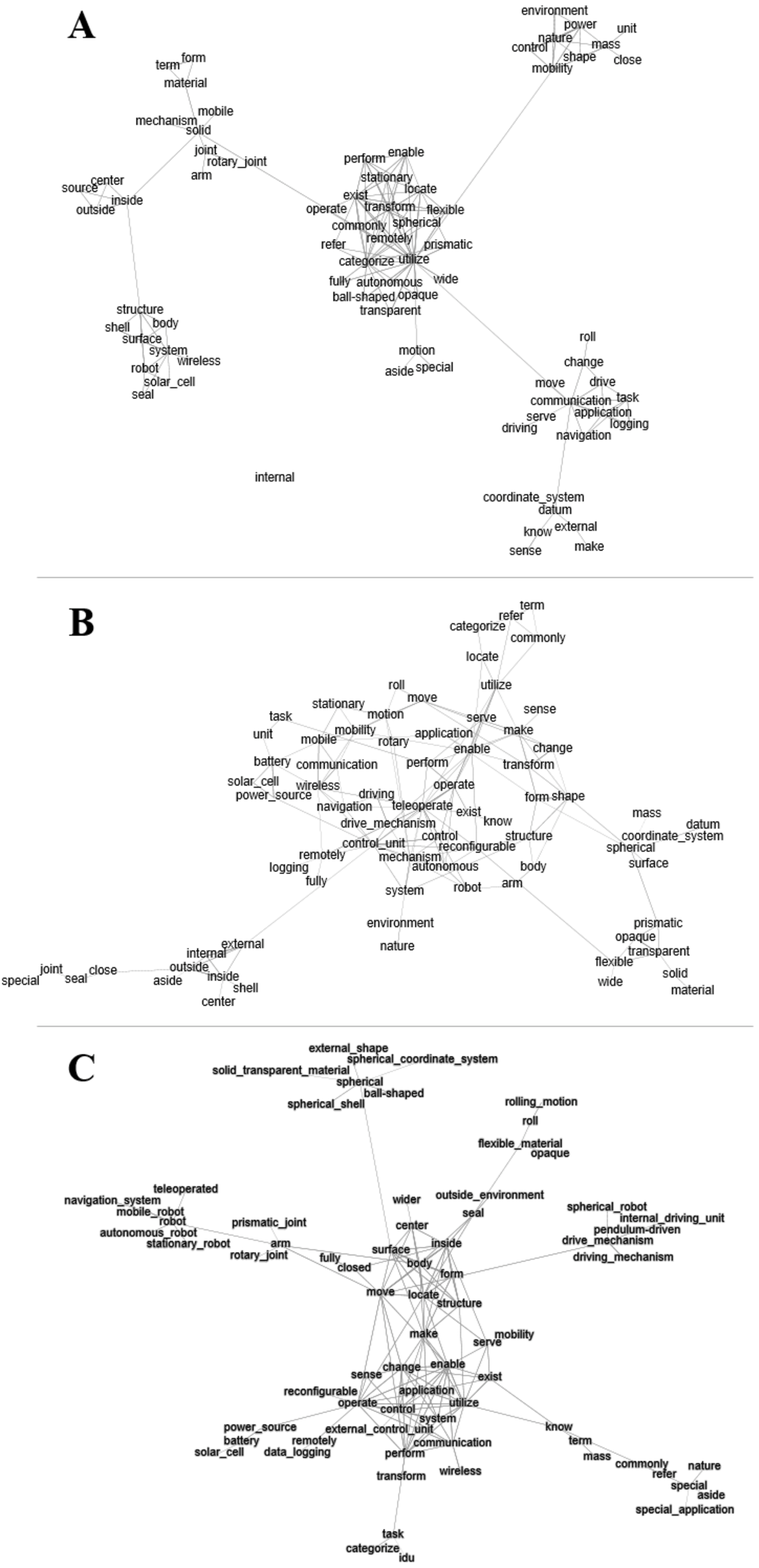}
	\caption{Graph visualizations based on A) WordNet, B) ConceptNet, and C) TechNet}
	\label{fig:fig3}
\end{figure}

The pre-processing step is followed by generating different graph representations of these retrieved terms for WordNet, ConceptNet, and TechNet (See Figure \ref{fig:fig3}). The graphical properties of the visualizations are adjusted as mundane and straightforward as possible to minimize any bias or effect that may be originated by choices such as color, font type, and size.

Among three different representations, TechNet (see Figure \ref{fig:fig3}C) confined general terms such as “body”, “form”, “locate”, and “move” in the center while creating cliques around the center that clump closely related components/design entities together. These clumped entities can be recombined to create the spherical robot concept described in the text. For example, the concepts and components related to the shape of the spherical robot are positioned very closely (spherical, ball-shaped, spherical shell), as do the ones related to electrical power. On the other hand, WordNet (Figure \ref{fig:fig3}A) groups some key aspects of spherical robots in the center of the graph, such as spherical, autonomous, ball-shaped. Unlike TechNet, WordNet lacks retrieving technology-specific phrases that contain essential information on technical design aspects and lacks distinctive categorization and cohesiveness in surrounding cliques. Lastly, ConceptNet (Figure \ref{fig:fig3}B) draws meaningful linkages in the center, but it lacks revealing distinct conceptually related cliques.

An online human participatory study was conducted to compare the effectiveness of these three knowledge bases in summarizing and representing the design-specific aspects of the technical definition via network mapping. By design aspects, the interrelations of the technical concepts and their spatial closeness due to interrelation strengths (i.e., the link weight in the graph, or semantic relevance in the context of knowledge bases) and grouping of them can be identified at first glance by most of the participants. For example, one participant may find it more representative to have closely related components and concepts in the same group, while another may prefer to see a direct link between the functions that he/she thinks as related.

The procedures of the study are as follows: 
\begin{enumerate}
	\item The participants are asked to read the technical description of the “spherical robot”. Then, they are asked to write a summary of the technical definition with a minimum of 50 words. This summarization task is included in the pilot study to force the participants to go through the technical definition thoroughly to understand the design details of the spherical robot. 
	\item The participants are asked to evaluate the randomly sorted graph representations about their performances of representing the specific design of the “spherical robot” concept. A 5-point Likert scale (“not representative”; “slightly representative”; “moderately representative”; “very representative”; “Strongly representative”) is presented to the participants for them to choose.
	\item The participants are finally asked which graph they consider the best, and why they consider it the best. This question is mainly used to resolve the situations in which a participant evaluates different representations with the same score.
\end{enumerate}

\begin{figure}[!h]
  \centering
  \includegraphics[scale=0.50]{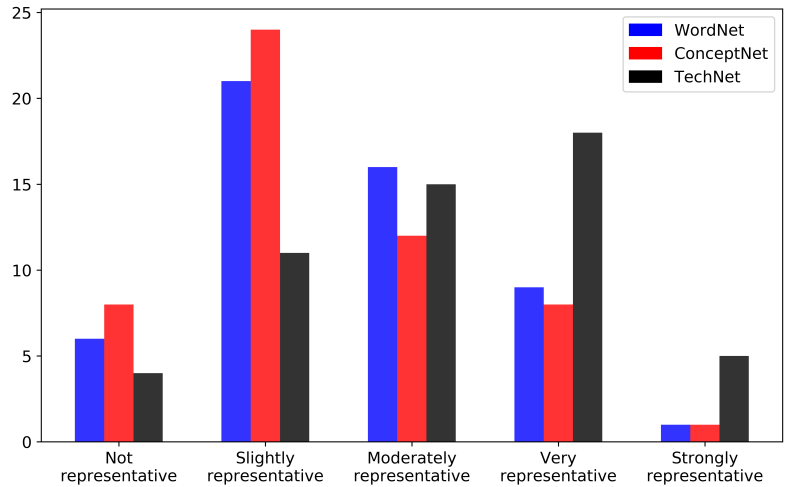}
  \caption{The distribution of participants' answers to evaluation questions.}
  \label{fig:fig4}
\end{figure}

The participants are selected among PhD students from SUTD (Singapore University of Technology and Design) and NTU (Nanyang Technological University) who have an engineering background and professional engineers actively working in the industry. No time limit was defined for individual sessions, but before starting, the participants are informed about the typical duration of the study, which is 10-15 minutes. In total, 56 participants completed the online pilot study. The first author of this study was able to perform informal interviews with 25 of them. Three participants among 25 indicated that they have got at least one question wrongly or comprehended it differently, or they have just chosen the first appeared graph representation since they were in a hurry. The data of these three participants were removed from the results. The rest of the participants evaluated the graph representations of the technical definition of “spherical robot” as presented in Figure \ref{fig:fig4}.

\begin{figure}[!h]
  \centering
  \includegraphics[scale=0.50]{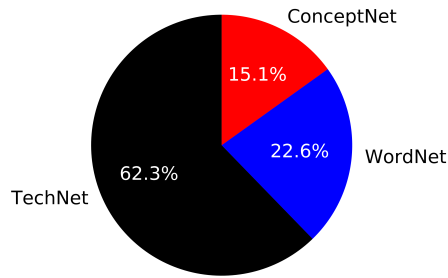}
  \caption{The percentage of participants according to their choice of best graph representation}
  \label{fig:fig5}
\end{figure}

Participants generally evaluated TechNet representation with higher scores comparing to others. While TechNet representation presents a left-skewed distribution, both WordNet and ConceptNet distributions' right skewness indicates their weaknesses in organizing and representing the technical information. Figure \ref{fig:fig5} aggregates the information given above using the participant’s response to the question asking the best of the three representations.

The main difference of the TechNet representation is that it can create distinct groups that are homogeneous within themselves, and together constitute sets of subsystems and/or abstract concepts that define the given technology. A few participants who chose TechNet as the best pointed out these specific characteristics. A participant commented, “The nodes of the graph include Multi-word Units (MWUs - e.g., spherical coordinate system) that are important for the reproduction of entities in the text description. The entities that comprise MWUs carry a specific meaning in this context, which is lost when these are decomposed into single word units”, which values the presence of particular keywords related to the given technology. Another participant pointed out the distinct groupings of subsystems with his/her comment: “Visually it splits my attention into a few key components which we associate as different parts that can be used to describe a spherical robot”.

WordNet leads to groups of terms, but it lacks to retrieve technology-specific terms that indicate design-specific clues for the given technology. A participant who favored WordNet indicated its advantage by commenting as “Because it is the most organized and simple to use and study” while another participant commented, “Arms are categorized in a more planned manner”.

On the other hand, sourcing ConceptNet relations result in a homogeneously drawn layout that lacks visibly distinctive groups. Still, some participants found the structure and the way ConceptNet connects the terms useful. For example, a participant evaluated the ConceptNet representation as the best and commented, “It has better connection between words as words which are related to each other are around each other. The other two graphs have multiple clusters which are connected to the center by a single connection which should not be right as all those words have relations with the words in the center”.

These participants' feedback and comments suggest that all three representations have some meaningful characteristics over others, but statistical results presented in Figures \ref{fig:fig4} and \ref{fig:fig5} strongly indicate TechNet’s superior performance for representing design-related information from a technical description.

\section{Discussion and Concluding Remarks}
\label{sec:conclusion}

Technical design text summarization is an important task, especially for prior art search and design learning. As reading and understanding complex technical documents require technical expertise and time, the concise representation of the information in an intuitive way can help designers easily, rapidly, and systematically understand the design described in the technical context. In this study, we focused on network graph visualization to abstract, represent and communicate the complex design information in technical texts, and exemplified the advantages of utilizing TechNet as the backend knowledge base for constructing the network, in comparison with the widely used common-sense knowledge bases in the engineering design research community, specifically WordNet and ConceptNet.

This study has various limitations. First, even though the introduced generic methodology for generating graph representations is based on a well-established force-directed layout and filtering methods, it is very specific. There are many other graph filtering and layout methods that can be employed and tested. However, covering various methodologies may bring difficulties in evaluating them in this setup, using human participants. Second, graphs or networks are not the only way to create visual summaries of a sample text. Engineering design researchers also used methods based on word clouds and hierarchical structures such as mind-maps to create intuitive design knowledge representations. Third, other possibly useful knowledge bases than the ones included in the present study can be explored. 

In sum, this study is only a first step in exploring abstract summary representations of complex technical documents. It should be viewed as an invitation for further research, methodological improvements, and applications that leverage TechNet and other engineering and technology-oriented knowledge bases to represent design-related knowledge, information and data.

\bibliographystyle{unsrt}  
\bibliography{references.bib}  

\end{document}